**Consequence of the intrachain dimer-monomer spin frustration and the interchain dimer-monomer spin exchange in the diamond-chain compound Azurite Cu$_3$(CO$_3$)$_2$(OH)$_2$**

J. Kang[1], C. Lee[1], R. K. Kremer[2] and M.-H. Whangbo[1]

[1] Department of Chemistry, North Carolina State University, Raleigh, North Carolina 27695-8204

[2] Max-Planck-Institut für Festkörperforschung, Stuttgart, Germany

Email: mike_whangbo@ncsu.edu

**Abstract**

The spin lattice appropriate for Azurite $Cu_3(CO_3)_2(OH)_2$ was determined by evaluating its spin exchange interactions on the basis of first principles density functional calculations. It is found that Azurite is not described by an isolated diamond chain with no spin frustration, but by a two-dimensional spin lattice in which diamond chains with spin frustration interact through the interchain spin exchange in the ab-plane. Our analysis indicates that the magnetic properties of Azurite at low temperatures can be approximated by two independent contributions, i.e., an isolated dimer and an effective uniform chain contributions. This prediction was verified by analyzing the magnetic susceptibility and specific heat data of Azurite.

## 1. Introduction

The interpretation of the magnetic data for a given magnetic solid begins with selecting a proper spin lattice and the associated spin Hamiltonian [1,2]. The spin lattice of a magnetic system is defined by the topology of the spin exchange paths one selects for the system, and its importance lies in the fact that the topology governs the nature of the magnetic energy spectrum and hence that of the magnetic properties. Ultimately, therefore, a correctly chosen spin lattice should be consistent with the electronic structure of the magnetic system because the latter determines the magnetic energy spectrum [1,2].

Experimentally, the spin exchange parameters of a selected spin lattice are determined as the fitting parameters that best reproduce the experimental data, typically, the spinwave dispersion relations from inelastic neutron scattering, the temperature-dependence of magnetic susceptibility or that of specific heat. When the observed magnetic data are not explained by a chosen spin lattice, one either attempts to improve it by introducing additional exchange parameters or searches for an alternative spin lattice. An unfortunate pitfall of such a fitting analysis is that more than one spin lattice may fit the same experimental data so that, even when a given spin lattice provides an excellent fitting, its correctness is not guaranteed, as found for $(VO)_2P_2O_7$ [3,4], $Na_3Cu_2SbO_6$ and $Na_2Cu_2TeO_6$ [5-9], and $Bi_4Cu_3V_2O_{14}$ [10-13], to name a few. Electronic structure calculations have proven to be extremely valuable and helpful in identifying the leading exchange parameters of a magnetic solid and hence in correctly identifying its spin lattice, which is not immediately apparent from geometrical-pattern considerations [2-13]. However, the choice of spin lattices is often guided by the geometrical pattern of the magnetic ion arrangement and/or the novelty of the physics the chosen model generates.

Recently, the diamond-chain model (see **Fig. 1**) has received much attention [10-20] due to the interesting theoretical questions associated with geometric spin frustration [21,22]. Because of the diamond-chain-like pattern of its $Cu^{2+}$ ion arrangement, $Bi_4Cu_3V_2O_{14}$ has been considered as a representative diamond-chain system [10-12] but it has been a puzzle that spin-frustration features expected for a diamond chain model are not present in $Bi_4Cu_3V_2O_{14}$ [10]. A recent electronic structure study showed that the correct spin lattice of $Bi_4Cu_3V_2O_{14}$ is not a diamond chain but an antiferromagnetic (AFM) chain made up of AFM linear trimers coupled through their midpoints [13]. The latter model predicts an AFM spin ground state with no spin frustration, in agreement with experiment. Another system actively probed in connection with the diamond-chain model is the mineral Azurite $Cu_3(CO_3)_2(OH)_2$ [14-20]. The magnetic susceptibility $\chi(T)$ of Azurite shows two broad peaks at ~22 K and ~4.4 K. Initially, Kikuchi *et al*. [14,15] interpreted the high-temperature part of the susceptibility of Azurite in terms of the diamond-chain model with spin frustration (i.e., AFM spin exchange $J_2$, $J_1$ and $J_3$ in Fig. 1). In explaining the low-temperature part of the susceptibility, namely, the double-peak feature of $\chi(T)$, it was found necessary [16,18] to employ the diamond chain model with no spin frustration [i.e., AFM $J_2$ and $J_1$, and ferromagnetic (FM) $J_3$]. More recently, Rule *et al.* [20] analyzed their specific heat and inelastic neutron scattering data in terms of the diamond-chain model without spin frustration by introducing two additional spin exchange parameters $J_m$ and $J_d$ (**Fig. 1**). Their fitting analysis led to the exchange parameters, $J_2/k_B = 55$ K, $J_1/k_B = 1$ K, $J_3/k_B = -20$ K, $J_m/k_B = 10.1$ K and $J_d/k_B = 1.8$ K. Given the structural parameters associated with these spin exchange paths (**Table 1**) [23,24] and the well-known structure-property relationships governing spin exchange interactions [2,25], the exchange parameters of Rule *et al*. raise the following questions:

(a) The Cu-O-Cu superexchange paths $J_1$ and $J_3$ are very similar. Namely, Cu…Cu = 3.275 Å and $\angle$Cu-O-Cu = 113.7° for $J_1$, and Cu…Cu = 3.290 Å and $\angle$Cu-O-Cu = 113.4° for $J_3$. Thus, it is unlikely that $J_1$ and $J_3$ can differ markedly in sign and magnitude.

(b) The $\angle$Cu-O-Cu angle for $J_3$ (113.4°) is much greater than 90°. Therefore, it is unlikely that the Cu-O-Cu superexchange $J_3$ can be strongly FM instead of being AFM.

(c) Adjacent $CuO_4$ monomers in each diamond chain have an arrangement leading to a negligible overlap between their magnetic orbitals. Thus, it is unlikely that the Cu-O…O-Cu super-superexchange $J_m$ can be as strongly AFM as reported by Rule *et al*. [20].

(d) The diamond-chain model proposed so far to analyze the magnetic properties of Azurite neglects the Cu-O…O-Cu super-superexchange $J_4$ between adjacent diamond chains in the ab-plane (**Fig. 1**). Because of the short O…O contact distance (2.219 Å) through a $CO_3$ bridge, this interchain interaction of the monomers of one chain with the dimers of its adjacent chains can be substantially AFM, thereby suggesting a two-dimensional (2D) character for Azurite. Thus, it is unlikely that a one-dimensional diamond-chain model is appropriate for Azurite. It is noted that, between adjacent diamond chains, one monomer interacts only with one dimer through the Cu-O…O-Cu super-superexchange $J_4$.

In the present work we probed the above four questions by evaluating the spin exchange interactions of Azurite on the basis of first principles DFT calculations and by analyzing the magnetic susceptibility and specific heat data of Azurite. Results of our calculations and analyses are presented in the following.

## 2. Calculations

Our calculations employed the Vienna ab-initio simulation package [26-28], the generalized gradient approximations (GGA) for the exchange and correlation corrections [29], the plane-wave cutoff energy of 400 eV, 196 k-points for the irreducible Brillouin zone, and the threshold of $10^{-6}$ eV for the self-consistent-field convergence of the total electronic energy. To properly describe the electron correlation of the Cu 3d states, the GGA plus on-site repulsion U (GGA+U) method [30] was employed with an effective U on the Cu atom. To check the dependence of our results on U, our analysis was carried out with U = 4 and 6 eV.

## 3. Results and discussion

To extract the values of the seven exchange parameters $J_1$, $J_2$, $J_3$, $J_4$, $J_m$, $J_d$, and $J_{d*}$, we perform GGA+U calculations for the nine ordered spin states depicted in **Fig. 2**, namely, six with antiferromagnetically coupled dimers and three with ferromagnetically coupled dimers. Note that there are two different spin exchange interactions between adjacent dimers, i.e., $J_d$ for AFM dimers and $J_{d*}$ for FM dimers. The relative energies of the nine ordered spin states obtained from GGA+U calculations are listed in **Table 2**. The total spin exchange interaction energies of the nine ordered spin states can be expressed in terms of the spin Hamiltonian,

$$\hat{H} = \sum_{i<j} J_{ij} \hat{S}_i \cdot \hat{S}_j,$$

where $J_{ij}$ is the spin exchange between the spin sites i and j, i.e., $J_{ij} = J_1, J_2, J_3, J_4, J_m, J_d$, or $J_{d*}$. By applying the energy expressions obtained for spin dimers with N unpaired spins per spin site (in the present case, N = 1) [31], the total spin exchange energies per two formula units are written as

$$E(A1) = (-2J_2 + 4J_1 - 4J_3 - 2J_m - 2J_d - 4J_4)N^2/4$$

$$E(A2) = (-2J_2 - 2J_m + 2J_d)N^2/4$$

$$E(A3) = (-2J_2 + 2J_m - 2J_d)N^2/4$$

$$E(A4) = (-2J_2 + 2J_m + 2J_d)N^2/4$$

$$E(A5) = (-2J_2 - 2J_1 + 2J_3 + 2J_4)N^2/4$$

$$E(A6) = (-2J_2 + 4J_1 - 4J_3 - 2J_m - 2J_d + 4J_4)N^2/4$$

$$E(F1) = (+2J_2 + 2J_m + 2J_{d*})N^2/4$$

$$E(F2) = (+2J_2 + 2J_m - 2J_{d*})N^2/4$$

$$E(F3) = (+2J_2 + 4J_1 + 4J_3 + 2J_m + 2J_{d*} + 4J_4)N^2/4$$

Thus, by mapping the relative energies of the nine ordered spin states determined from the GGA+U calculations onto the corresponding energies obtained from the total spin exchange energies, we find the values of $J_2$, $J_1$, $J_3$, $J_m$, $J_d$, $J_{d*}$ and $J_4$ summarized in **Table 3**. The GGA+U calculations with both U = 4 and 6 eV provide the same trends in the relative strengths of the spin exchange parameters. $J_m$, $J_d$ and $J_{d*}$ are negligibly weak compared with $J_2$, $J_1$, $J_3$ and $J_4$. The values of $J_2$, $J_1$, $J_3$ and $J_4$ obtained with U = 4 eV are slightly larger than those obtained with U = 6 eV, which is understandable because the magnitude of an AFM spin exchange is inversely proportional to U [2]. As generally observed for GGA+U calculations [8,13,31], these values overestimate the exchange parameters. The spin exchange parameters calculated with U = 4 eV are overestimated by a factor of ~4 if our $J_2$ value is compared with that found by Rule *et al*.

As anticipated, the calculated $J_1$ and $J_3$ are both AFM and are similar in magnitude ($J_1/J_2 = 0.25$ and $J_3/J_2 = 0.24$ with U = 4 eV; $J_1/J_2 = 0.24$ and $J_3/J_2 = 0.21$ with U = 6 eV), $J_m$ and $J_d$ are very weak ($J_m/J_2 = 0.00$, $J_d/J_2 = -0.02$ with U = 4 eV; $J_m/J_2 = 0.01$, $J_d/J_2 = 0.00$ with U = 6 eV), and the interchain exchange $J_4$ is substantially AFM ($J_4/J_2 = 0.13$ with U = 4 and 6 eV) and comparable to the intrachain exchange $J_1$ and $J_3$. Thus, the diamond chains are spin frustrated as initially suggested [14,15], but the interchain interaction $J_4$ in the *ab*-plane is substantial leading to a 2D spin lattice model for Azurite. Thus, our study answers affirmatively for all four questions raised in Introduction.

It is important to consider a simple spin lattice model for Azurite that captures the essence of its exchange interactions $J_2$, $J_1$, $J_3$ and $J_4$. Since $J_2 \gg J_1$, $J_3$, $J_4$, an isolated dimer model would be reasonable for Azurite at high temperatures. At low temperatures, the dimer-monomer exchanges $J_1$, $J_3$ and $J_4$ cannot be neglected. Within each diamond chain, the dimer-monomer exchanges are frustrated with $J_1 \approx J_3$. Consequently, at low temperatures, the interchain dimer-monomer exchange $J_4$ becomes more important than the intrachain dimer-monomer exchanges $J_1$ and $J_3$. Across the diamond chains, the exchange paths $J_4$ and $J_2$ form $(-J_4-J_4-J_2-)_\infty$ chains. In each $(-J_4-J_4-J_2-)_\infty$ chain the monomers interact through the dimers, which are in singlet state at low temperatures since $J_2 \gg J_4$. This implies that the monomers of each $(-J_4-J_4-J_2-)_\infty$ chain would effectively behave as if they formed a uniform chain with spin exchange $J_4$. To a first approximation, therefore, the magnetic susceptibility and the specific heat data of Azurite can be approximated in terms of two independent contributions, i.e., an isolated dimer (described by $J_{dimer}$) and a uniform chain (described by $J_{chain}$) contributions. Thus, we used this model to fit the magnetic susceptibility data of Azurite given by Kikuchi *et al.*[15], who determined the susceptibilities with the magnetic field aligned along the directions parallel and perpendicular to the b-axis (*H*∥b and *H*⊥b, respectively). In our fitting analysis, the temperature-independent contributions to the susceptibility from core diamagnetic and Van Vleck paramagnetic contributions were estimated to cancel out (see, e.g., the discussion in [32]) and hence were not taken into account in the fitting. As shown in **Fig. 3a**, the results of the fitting is excellent, and the ratio $J_{dimer}/J_{chain}$ = 0.15 obtained from the fitting is very close to the ratio of $J_2/J_4$ = 0.13. The anisotropy of the $H\|b$ and $H\perp b$ magnetic susceptibilities can be related to the g-factor anisotropy of the Cu moments in the dimers and the monomers (see the caption of **Fig. 3a** for the fitted g-factors).

To provide another test for our conclusion, we carried out a fitting analysis for the specific heat of Azurite in a similar manner. Although the specific heat of Azurite has been reported in several studies [e.g., 15, 20], we have re-determined it down to 0.4 K using a PPMS calorimeter with a $^3$He extension (Quantum Design, 6325 Lusk Boulevard, San Diego, CA). To subtract the lattice contribution to our specific heat data, we fitted the coefficients $a_i$ of an odd-polynomial, $\Sigma_i\, a_i T^{2i+1}$ ($i$ = 1, 2, …, 6), to the high temperature heat capacity data. To account for the decreasing magnetic contributions to the specific heat at high temperatures, we also added a term proportional to $1/T^2$. **Fig. 3b** shows the magnetic contributions to the specific heat capacity (hereafter the magnetic specific heat, $C_{mag}$) of Azurite. As shown by the solid curves of **Fig. 3b**, the characteristic features and the temperature dependence of the magnetic specific heat are very well reproduced by our model. The resulting $J_{dimer}$ and $J_{chain}$ values are very close to the corresponding values obtained from the susceptibility fitting, and the ratio $J_{chain}/J_{dimer}$ = 0.13 is the same as the ratio $J_4/J_2$. The small peak of the specific heat around ~1.9 K (**Fig. 3b**) is due to a long-range magnetic ordering.

In a recent NMR study [33], Aimo *et al*. suggested that the diamond chain model may be too simple to describe the magnetic properties of Azurite, and the possibility of interchain couplings. These suggestions are consistent with our conclusions.

## 4. Concluding remarks

In summary, the diamond-chain spin lattice with no spin frustration employed for the description of the magnetic properties of Azurite is not consistent with the electronic structure of Azurite. Our calculations show that Azurite should be described by the 2D spin lattice in which diamond chains with spin frustration are coupled through the interchain exchange $J_4$ between monomers and dimers. Due to the spin frustration associated with $J_1$ and $J_3$ as well as the fact that $J_2 >> J_4$, the magnetic susceptibility and the specific heat of Azurite are well approximated in terms of the two independent contributions, i.e., the isolated dimer (defined by $J_2$) and the effective uniform chain (defined by $J_4$) contributions. Our work shows the importance of choosing spin lattices on the basis of electronic structure considerations [2]. Interesting but erroneous interpretations often result as found for Azurite and other systems [3-13], when the choice is made by inspecting the geometrical pattern of the magnetic ion arrangement or by seeking the novelty of the physics the chosen model generates.

**Acknowledgment**

This work was supported by the Office of Basic Energy Sciences, Division of Materials Sciences, U. S. Department of Energy, under Grant DE-FG02-86ER45259 and by the resources of the NERSC Center and the HPC Center of NCSU.

Table 1. Geometrical parameters associated with the spin exchange paths of Azurite $Cu_3(CO_3)_2(OH)_2$

|  | Cu…Cu | $\angle$Cu-O-Cu | O…O |
|---|---|---|---|
| $J_2$ | 2.983 | 97.9 | - |
| $J_1$ | 3.275 | 113.7 | - |
| $J_3$ | 3.290 | 113.4 | - |
| $J_m$ | 5.849 | - | 2.597 |
| $J_d$, $J_{d*}$ | 5.849 | - | 3.893 |
| $J_4$ | 4.872 | - | 2.219 |

[a] The Cu…Cu and O…O distances are in units of Å, and the $\angle$Cu-O-Cu angles in units of degrees.

Table 2. Relative energies (meV per two formula units) of the nine ordered spin states

obtained from GGA+U calculations with U = 4 and 6 eV

| U | A1 | A1 | A3 | A4 | A5 | A6 | F1 | F2 | F3 |
|------|------|------|------|------|------|------|-------|-------|-------|
| 4 eV | 0.00 | 4.45 | 5.04 | 4.46 | 6.60 | 7.98 | 35.93 | 36.20 | 55.04 |
| 6 eV | 0.00 | 1.41 | 1.50 | 1.51 | 2.38 | 4.81 | 20.35 | 20.88 | 31.28 |

Table 3. Spin exchange parameters (in units of $k_BK$) of Azurite $Cu_3(CO_3)_2(OH)_2$ determined by GGA+U calculations with U = 4 and 6 eV [a]

|  | U = 4 eV | U = 6 eV |
|---|---|---|
| $J_2$ | 363.3  (1.00) | 221.7  (1.00) |
| $J_1$ | 89.4  (0.25) | 52.6  (0.24) |
| $J_3$ | 86.1  (0.24) | 46.3  (0.21) |
| $J_m$ | 0.1  (0.00) | 1.2  (0.01) |
| $J_d$ | -6.7  (-0.02) | 0.15  (0.00) |
| $J_{d*}$ | -3.0  (-0.01) | -6.1  (-0.03) |
| $J_4$ | 46.3  (0.13) | 27.9  (0.13) |

[a] The numbers in the parentheses are relative numbers with respect to $J_2$.

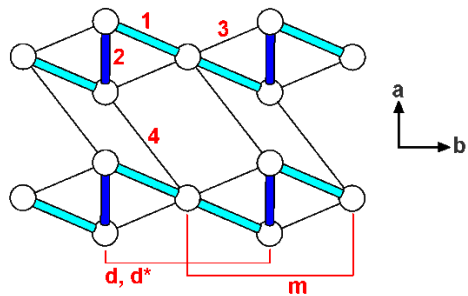

Figure 1.    Spin exchange paths of Azurite shown with two diamond chains in the ab-plane, where the labels 1, 2, 3, 4, m, d and d* refer to the spin exchange paths $J_1$, $J_2$, $J_3$, $J_4$, $J_m$ and $J_d$, and $J_{d*}$, respectively. See text for the difference between $J_d$ and $J_{d*}$.

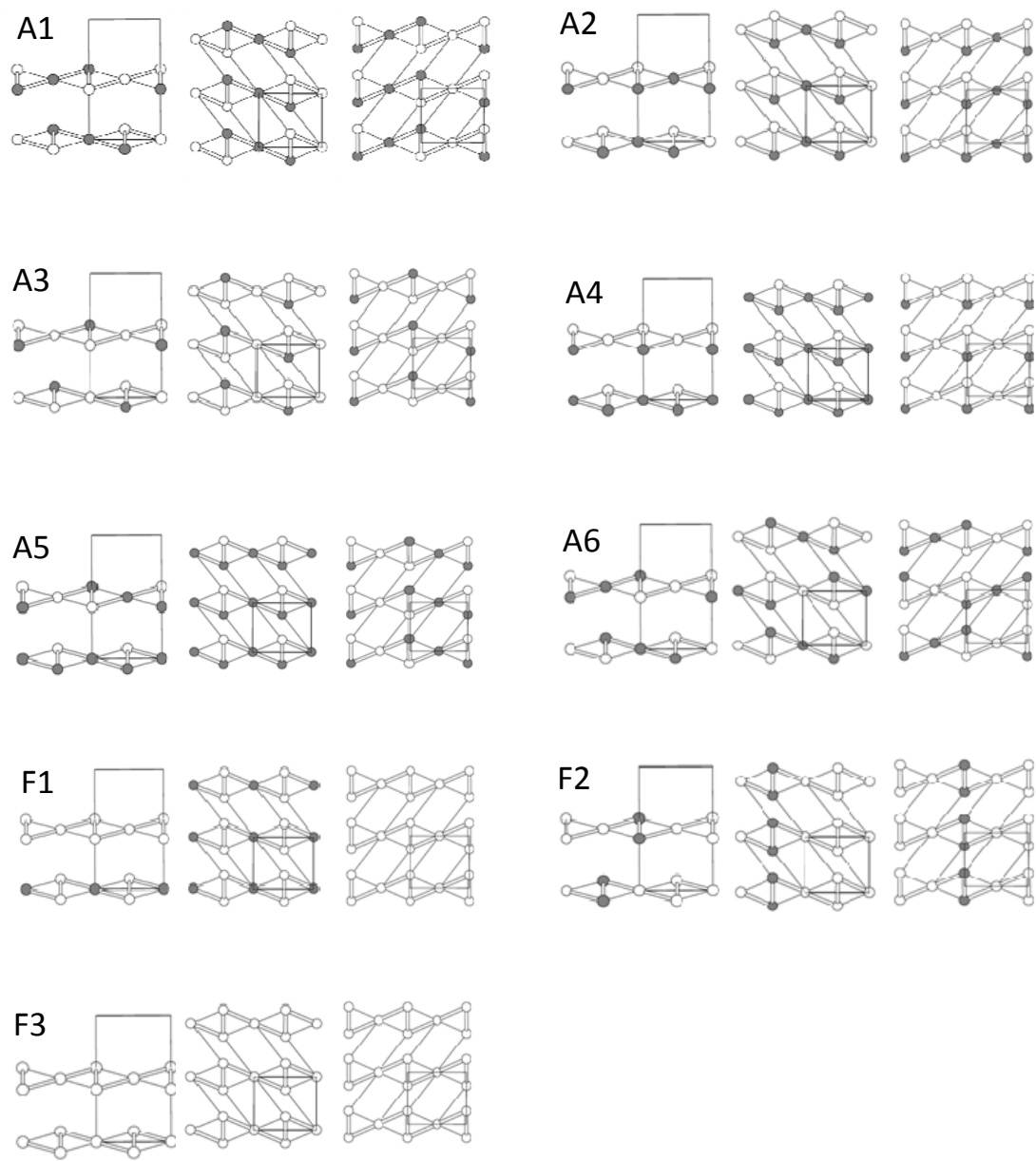

Figure 2.      Nine ordered spin states of Azurite used for GGA+U calculations to extract the spin exchange parameters $J_1$, $J_2$, $J_3$, $J_4$, $J_m$ and $J_d$, and $J_{d*}$. See text for the difference between $J_d$ and $J_{d*}$. Up and down spins at the Cu sites are indicated by shaded and unshaded circles, respectively. For each ordered spin state, the left diagram shows a bc-plane projection view of two diamond-chains. The middle and right diagrams depict how the diamond-chains repeat in the two different layers parallel to the ab-plane.

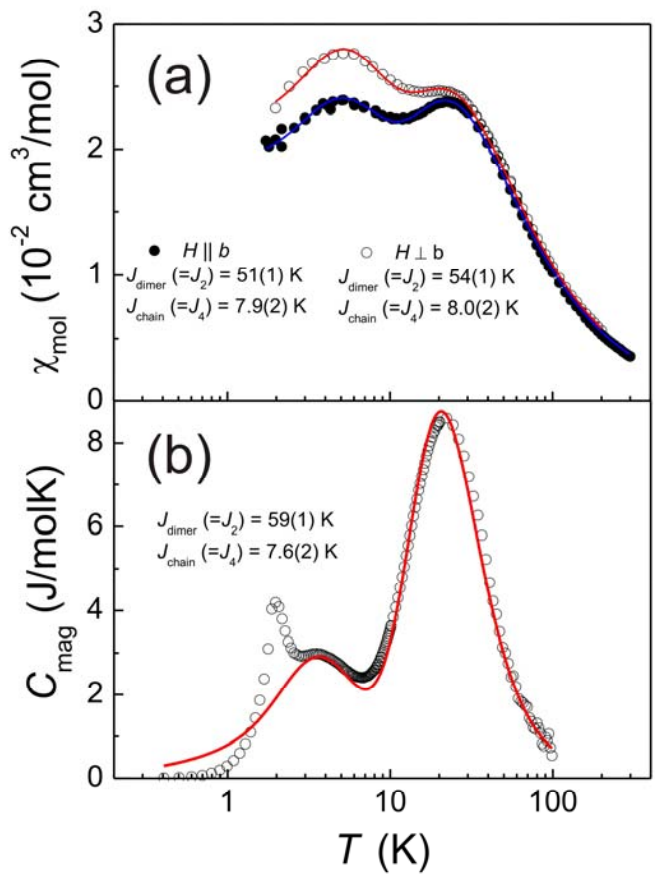

Figure 3. (a) Magnetic susceptibility data of Azurite determined by Kikuchi *et al*. [15] with the magnetic field applied along the direction parallel and perpendicular to the crystallographic c-direction: $H||b$ (●) and $H\perp b$ (○). The solid lines represent the fitted curves in terms of the $S = 1/2$ AFM dimer (with $J_{dimer}$) and the $S = 1/2$ Heisenberg uniformAFM chain (with $J_{chain}$) contributions. The fitted $J_{dimer}$ and $J_{chain}$ values are given in the figure. The anisotropy of the magnetic susceptibility is well accounted for by using different g-factors for the monomer (constituting the chain) and the dimer spins. For $H||b$, the g-factors are 1.86 and 2.14 for the monomer and the dimer, respectively. For $H\perp$b, the g-factors are 2.02 and 2.12 for the monomer and dimer spins, respectively. (b) Magnetic contribution to the specific heat of Azurite, $C_{mag}$, obtained as described in the text. The solid curve represents the fitting in terms of the $S = 1/2$ AFM dimer (with $J_{dimer}$) and the $S = 1/2$ Heisenberg uniform AFM chain (with $J_{chain}$) contributions. The fitted exchange parameters are shown in the figure.